\documentclass[aps,prl,superscriptaddress]{revtex4}
\usepackage{amsmath}
\usepackage{graphicx}

\begin{document}

\title{Spin Transfer Torques induced by Spin Hall Effect}

\author{A. Vedyayev}
\affiliation{SPINTEC, UMR 8191 CEA-INAC/CNRS/UJF-Grenoble 1/Grenoble-INP, 38054 Grenoble, France}
\affiliation{Department of Physics, Moscow Lomonosov State University, Moscow 119991, Russia}
\author{N. Strelkov}
\affiliation{SPINTEC, UMR 8191 CEA-INAC/CNRS/UJF-Grenoble 1/Grenoble-INP, 38054 Grenoble, France}
\affiliation{Department of Physics, Moscow Lomonosov State University, Moscow 119991, Russia}
\author{M. Chshiev}
\affiliation{SPINTEC, UMR 8191 CEA-INAC/CNRS/UJF-Grenoble 1/Grenoble-INP, 38054 Grenoble, France}
\author{N. Ryzhanova}
\affiliation{SPINTEC, UMR 8191 CEA-INAC/CNRS/UJF-Grenoble 1/Grenoble-INP, 38054 Grenoble, France}
\affiliation{Department of Physics, Moscow Lomonosov State University, Moscow 119991, Russia}
\author{B. Dieny}
\affiliation{SPINTEC, UMR 8191 CEA-INAC/CNRS/UJF-Grenoble 1/Grenoble-INP, 38054 Grenoble, France}
\email{vedy@magn.ru}

\begin{abstract}
Spin accumulation and spin transfer torques induced by Spin Hall Effect in bi-layer structures comprising ferromagnetic and paramagnetic materials are theoretically investigated. The charge and spin diffusion equations taking into account spin-flip and spin Hall effect are formulated and solved analytically and numerically
for in structures with simplified and complex geometry, respectively. It is demonstrated that spin torques could be efficiently produced by means of Spin Hall effect which may be further enhanced by modifying structure geometry.
\end{abstract}

\maketitle

\section{Introduction}
In~\cite{Dyakonov_Perel_71_1,Dyakonov_Perel_71_2}
Dyakonov and Perel predicted the existence of Spin Hall Effect (SHE)
in paramagnetic metal without applying the external magnetic field under
influence of spin-orbit interaction. It was found later a lot of similarities
between the ``anomalous'' Hall Effect in ferromagnetic metals (FM) and SHE~\cite{Fert1981231,Hirsch_99},
assuming that in both cases the mechanisms of creation of the transversal
electric current is the skew scattering or side jump mechanism due to the
spin-orbit interaction. The more detailed analyses of SHE was presented
in~\cite{Zhang_SH_2000, Shchelushkin_Brataas_2005_1, Shchelushkin_Brataas_2005_2}
employing the semi-classical Boltzmann equation
in~\cite{Zhang_SH_2000} and Keldish formalism
in~\cite{Shchelushkin_Brataas_2005_1, Shchelushkin_Brataas_2005_2}
for the calculation of the transport properties of the paramagnetic
metal taking into account the spin-orbit interaction. It is considered that SHE may
be very effective tool for the manipulation with spin current and spin accumulation.
The most interesting could be the hybrid structure consisting of ferromagnetic (FM) 
metal/paramagnetic (PM) metal with large SHE, if both mechanisms of
creation of the spin current are involved: SHE and spin polarization by ferromagnetic metal. 
In this work we investigate spin transfer torques produced by the interplay between these mechanisms
in FM/NM bi-layer structures by solving charge and spin diffusion equations.

\section{Model}
For the calculation both spin accumulation and spin polarized current in such a
structure which may have a complex geometry, it is necessary to derive
spin diffusion equation taking into account both SHE for the paramagnetic
metal and processes governing the spin transport in ferromagnetic metal.
It is necessary also to develop the convenient code for the numerical
simulations for the system of the complicated geometry in 2D and 3D cases.
Following~\cite{Valet_Fert_1993, Zhang_Levy_Fert_02,
Strelkov_IEEE_10,Strelkov_PRB_11} where the diffusion equation describing spin transport
in ferromagnetic multilayered structures where developed and with diffusion
equation obtained in~\cite{Shchelushkin_Brataas_2005_2}
and describing SHE we can write down:

\begin{equation}
\vec j_e=-\sigma_0\vec\nabla\varphi
-\beta\frac{\sigma_0}{e\nu}\vec\nabla(\vec U_M, \vec m)
+a_0^3\sigma_{\mathrm{SH}}\left[ \vec m \times \vec\nabla\varphi \right]
\label{eq:je}
\end{equation}

\begin{equation}
\vec j_m^{(i)}=-\beta\sigma_0\vec\nabla\varphi U_M^{(i)}
-\frac{\sigma_0}{e\nu}\vec\nabla m^{(i)}
-\sigma_{\mathrm{SH}}U_m^{(i)}\left[ \vec U_m \times \vec\nabla\varphi \right]
\label{eq:jm}
\end{equation}

\begin{equation}
\begin{cases}
  \mathrm{div}\vec j_e=0 \\
  \mathrm{div}\vec j_m\,^{(i)}=
     -\frac{\sigma_0}{e^2\nu l_J^2}\left[ \vec m \times \vec U_M \right]^{(i)}
     -\frac{\sigma_0}{e^2\nu l_{\mathrm{sf}}^2}m^{(i)},
\end{cases}
\label{eq:eq}
\end{equation}
where $\sigma_0$ is the conductivity, $\beta$ is the spin-asymmetry parameter of conductivity,
$\sigma_{\mathrm{SH}}$ is the spin Hall conductivity, $\nu$ -- density of states,
$\vec U_M=\vec M/M_s$, where $\vec M$ is magnetization vector in ferromagnetic,
$\vec U_M=\vec m/|\vec m|$, where $\vec m$  is spin accumulation vector,
index $i$ is a component of vectors $\vec m$, $\vec j_m$ and $\vec U_M$ in spin space,
$l_{\mathrm{sf}}$ -- spin diffusion length and $l_J$ -- exchange spin diffusion length.
Inside the ferromagnetic metal one have to omit the terms of SH, and in paramagnetic metal
$\beta=0$. In equation~(\ref{eq:je}) we added the last term,
which is quadratic on the $\vec\nabla\varphi$ as the value of $m$ is proportional to
$\vec\nabla\varphi$, and in~(\ref{eq:jm}) we omitted terms corresponding to the contribution
of the anomalous velocity (see (2) in~\cite{Shchelushkin_Brataas_2005_2}).
Here we have to mention the article~\cite{Fert_Levy_2011}, where it was proven
that the large $\sigma_{\mathrm{SH}}/\sigma_0$, experimentally observed for $Au$ doped by
$Fe$ and $Pt$ impurities~\cite{Takeshi_nmat_08,Takeshi_private_1}
and for $Cu$ doped by $Ir$~\cite{Niimi_11} may be attributed to the resonant
electron scattering on impurities if take into account spin-orbit interaction.

Let us consider the system, consisting of two flat layers, one of the paramagnetic
metal with SHE and second ferromagnetic layer with current in $x$ direction.
If to consider the case $L_x>L_y\gg L_z^F+L_z^P$, where $L_x$ and
$L_z^F+L_z^P$ are the lengths of the system in $x$ and $z$ directions,
the solution of (\ref{eq:eq}) in the region of $x$ far from
$(-L_x,L_x)$ may be easily found, and expression for $\varphi$,
$m^{(2)}$ ($m^{(0)}=m^{(3)}=0$) are the following:

\begin{equation}
m^{(1)}=-V\frac{e\nu}{\mathrm D}\frac{\sigma_{\mathrm{SH}}}{\sigma_1}
\frac{l_{\mathrm{sf},1}}{L_x}
\left[
\sinh\frac{L_1}{2l_{\mathrm{sf},1}}\sinh\frac{L_1+2z}{2l_{\mathrm{sf},1}}+
\frac{\sigma_2}{\sigma_1}\frac{l_{\mathrm{sf},1}}{l_{\mathrm{sf},2}}
\left(1-\beta^2\right)\tanh\frac{L_2}{l_{\mathrm{sf},2}}
\sinh\frac{z}{l_{\mathrm{sf},2}}
\right]
\end{equation}

\begin{equation}
m^{(2)}=-V\frac{e\nu}{\mathrm D}\frac{\sigma_{\mathrm{SH}}}{\sigma_1}
\frac{l_{\mathrm{sf},1}}{L_x}
\sinh^2\frac{L_1}{2l_{\mathrm{sf},2}}
\cosh\frac{L_2-z}{l_{\mathrm{sf},2}}/
\cosh\frac{L_2}{l_{\mathrm{sf},2}}
\end{equation}

\begin{eqnarray}
\lefteqn{
\varphi_1=\frac V2\left(1+\frac{x}{L_x}\right)-
V^2\frac{(e\nu)^2}{2 \mathrm{D}}\left(\frac{\sigma_{\mathrm{SH}}}{\sigma_1}
\frac{l_{\mathrm{sf},1}}{L_x}\right)^2\sinh^2\frac{L_1}{2l_{\mathrm{sf},1}}
}\nonumber\\
& & \times\left[
2\sinh\frac{L_1}{2l_{\mathrm{sf,1}}}\cosh\frac{L_1+2z}{2l_{\mathrm{sf},1}}+
\frac{\sigma_2}{\sigma_1}\frac{l_{\mathrm{sf},1}}{l_{\mathrm{sf},2}}
\frac
{\cosh\frac{z}{l_{\mathrm{sf},2}}-\cosh\frac{L_x}{l_{\mathrm{sf},1}}}
{1-\cosh\frac{L_x}{l_{\mathrm{sf},1}}}
\tanh\frac{L_2}{l_{\mathrm{sf},2}}
\right]
\end{eqnarray}

\begin{eqnarray}
\lefteqn{
\varphi_2=\frac V2\left(1+\frac{x}{L_x}\right)-
V^2\frac{(e\nu)^2}{2 \mathrm{D}}\left(\frac{\sigma_{\mathrm{SH}}}{\sigma_1}
\frac{l_{\mathrm{sf},1}}{L_x}\right)^2\sinh^2\frac{L_1}{2l_{\mathrm{sf},1}}
}\nonumber\\
& & \times\left[
\sinh\frac{L_1}{l_{\mathrm{sf,1}}}+
\frac{\sigma_2}{\sigma_1}\frac{l_{\mathrm{sf},1}}{l_{\mathrm{sf},2}}
\tanh\frac{L_2}{l_{\mathrm{sf},2}}
\right]-
V\left[\mathrm{sign}\,U_M^{(2)}\right]\frac{\beta}{\mathrm{D}}
\frac{\sigma_{\mathrm{SH}}}{\sigma_1}
\frac{l_{\mathrm{sf},1}}{L_x}
\frac
{\cosh\frac{L_2}{l_{\mathrm{sf},2}}-\cosh\frac{L_2-z}{l_{\mathrm{sf},2}}}
{\cosh\frac{L_2}{l_{\mathrm{sf},2}}}
\end{eqnarray}

\begin{equation}
D=\sinh\frac{L_1}{l_{\mathrm{sf},1}}+
\frac{\sigma_2}{\sigma_1}\frac{l_{\mathrm{sf},1}}{l_{\mathrm{sf},2}}
\left(1-\beta^2\right)\tanh\frac{L_2}{l_{\mathrm{sf},2}}
\cosh\frac{L_1}{l_{\mathrm{sf},1}},
\nonumber
\end{equation}
where $L_1$ and $L_2$ -- are thicknesses of SH-layer and FM-layer respectively.

\begin{figure}[ht]
\includegraphics[width=0.5\textwidth]{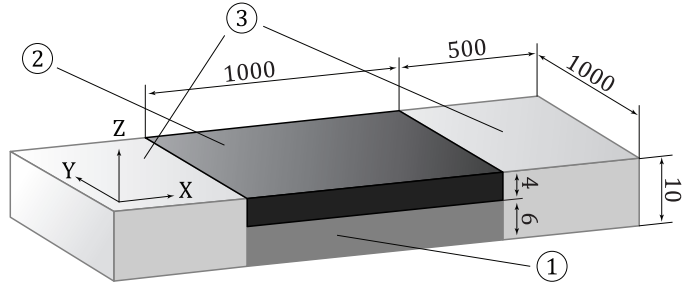} 
\caption{\label{fig:sys1000x1000} Schematic of $Pt$/$Py$ bylayer. Sizes are in ``nm''. 1 -- $Pt$ layer,
2 -- $Py$ layer, 3 -- $Cu$ electrodes. Current is along $x$ axe. Magnetisation of $Py$ is in
$xy$ plane at $\pi/4$ angle to $x$ axe.}
\end{figure}

\begin{figure}[ht]
\includegraphics[width=0.5\textwidth]{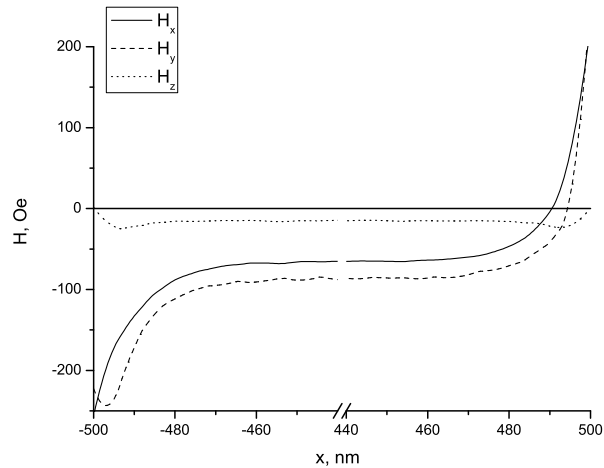} 
\caption{\label{fig:Hm1000} Effective fields induced by $Pt$ SHE in $Py$ near the
interface. Adopted values of the parameters: $\sigma^{Pt}=0.005 (\Omega\cdot nm)^{-1}$,
$l_{\mathrm{sf}}^{Pt}=10nm$, $\sigma_{\mathrm{SH}}^{Pt}=0.1\sigma^{Pt}$,
$\sigma^{Py}=0.0022 (\Omega\cdot nm)^{-1}$, $l_{\mathrm{sf}}^{Py}=6nm$,
$\beta=0.7$, $l_J=1nm$, current density $j=10^7A/cm^2$.}
\end{figure}

\begin{figure}[ht]
\includegraphics[width=0.7\textwidth]{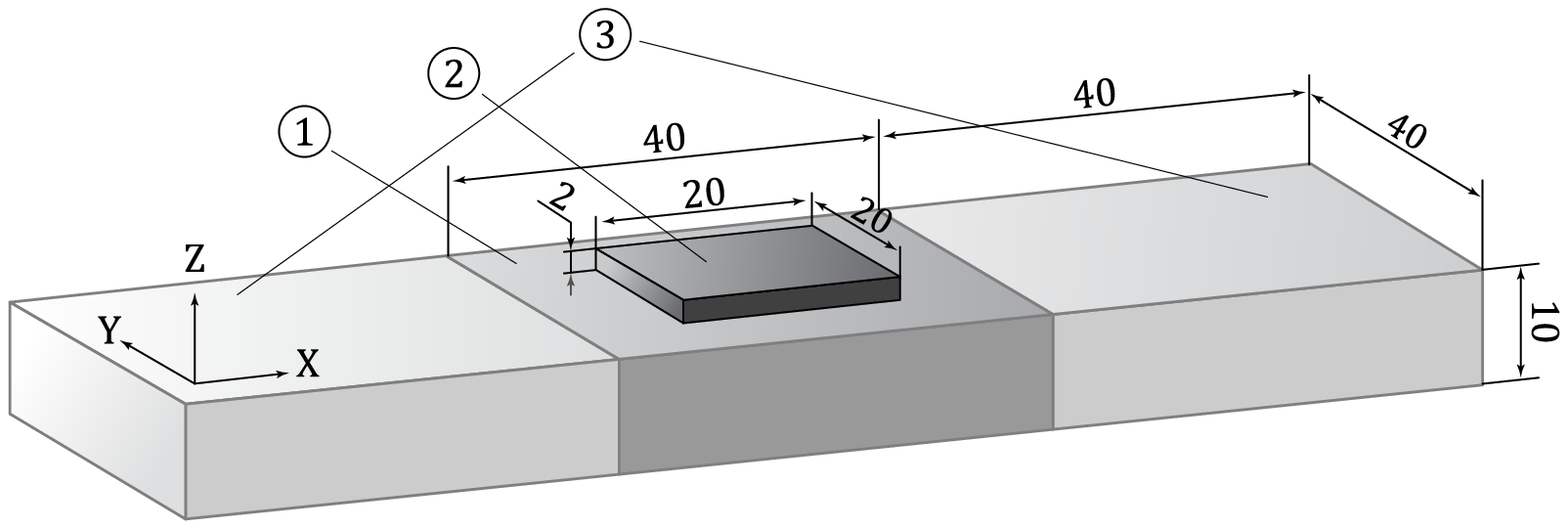} 
\caption{\label{fig:sys20x20} Schematic of $Pt$/$Py$ bylayer. Sizes are in ``nm''. 1 -- $Pt$ layer,
2 -- $Py$ layer, 3 -- $Cu$ electrodes. Current is along $x$ axe. Magnetisation of $Py$ is in
$xy$ plane at $\pi/4$ angle to $x$ axe or along $z$ axe.}
\end{figure} 

\begin{figure}[ht]
\includegraphics[width=0.5\textwidth]{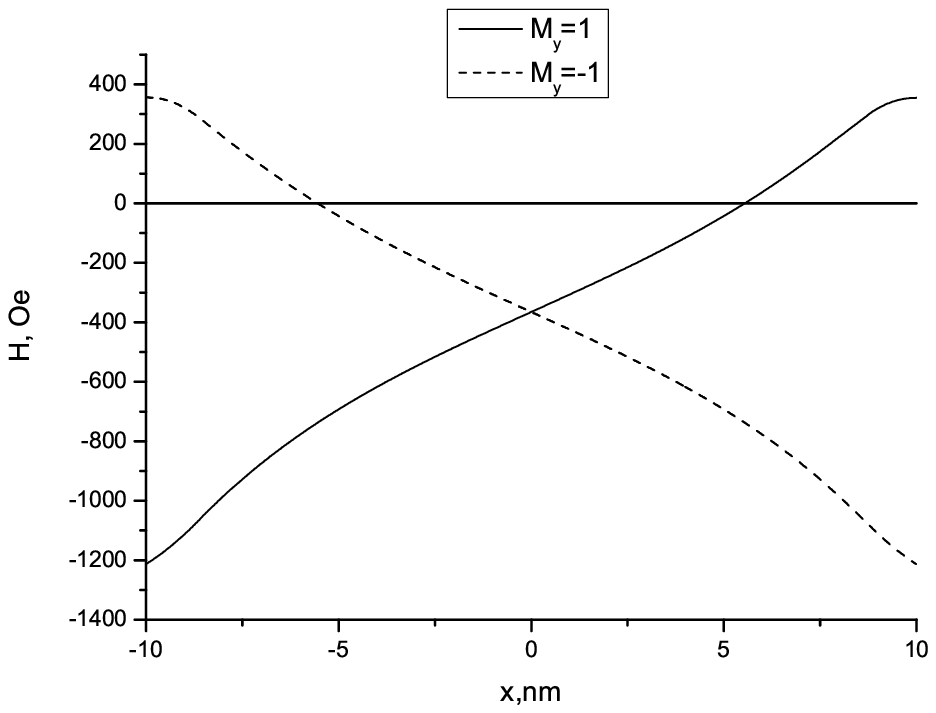} 
\caption{\label{fig:Hm20My} Effective fields in $Co$/$Pt$ structure for the direction of magnetisation
parallel and antiparallel to $y$ axe. Only $H_y$ component survive.
$\sigma^{Co}=0.005 (\Omega\cdot nm)^{-1}$, $l_{\mathrm{sf}}^{Co}=10nm$.}
\end{figure}
The spin accumulation $m^{(2)}$ produces the effective field $H_{\mathrm{eff}}$ acting on the magnetisation
of the ferromagnetic layer, which value is equal $H_{\mathrm{eff}}=m^{(2)}J_{sd}/\mu_B$, where $J_{sd}$
is $s$-$d$ exchange integral. It is important to notice that this field is proportional to drop of voltage and not
to current density, as in the case of Oersted field created by the current. So if one choose as a source of SHE
dirty paramagnetic metal only due to its higher resistance the value of the induced by SHE effective field for
the constant current density will increase, and besides that the following conclusion of~\cite{Fert_Levy_2011}
the value of $\sigma_{\mathrm{SH}}/\sigma_0$ may increase as well. Another interesting conclusion,
following from expression for the potential $\varphi$, is that SHE produce drop of voltage in $z$ direction
perpendicular to the current. If there is no ferromagnetic layer above SHE structure the drop of voltage is
symmetric ($\propto \cosh[z/l_{\mathrm{sf}}]$) and quadratic on the applied voltage $V$, but in
presence of FM layer this drop has linear on $V$ part and is finite across the thickness of the system. To
investigate the influence of the edge of layers on spin accumulation in the system of finite size we have solved
equation of~(\ref{eq:eq}) numerically using Comsol Multiphysics for several artificial system,
consisting of SHE substrate and ferromagnetic layers or dots situated on the top of this substrate.

\section{Results}

In Fig.\ref{fig:sys1000x1000} we give the schematic of $Pt$/$Py$ bilayer, which was experimentally investigated
in~\cite{Liu_Moriyama_11}. In fig.\ref{fig:Hm1000} the dependence on $x$ coordinate of the induced by SHE fields
$H_{\mathrm{SHE}}$ acting on the magnetization inside of $Py$ layer and near the $Pt$/$Py$
interface is shown. It is clear that SHE spin current produce all tree components of the field,
meanwhile Oersted field $H_j$ produced by the current in given geometry has only $y$-component,
and $H_j=4$ Oe. The $z$-component of the SHE field produces torque in the plane of the $Py$ layer,
and the ratio $H_{\mathrm{SHE}}/H_j$ lies within the interval $1.3\div3.7$, where $1.3$ is the average
value over the volume of $Py$ layer while $3.7$ represents this ratio near the interface, what is close
to the value recalculated using value of $S/A=0.63$ in~\cite{Liu_Moriyama_11} which gives 
$H_{\mathrm{SHE}}/H_j=\sqrt{1+4\pi M_{\mathrm{eff}}/H_{\mathrm{eff}}}\,S/A=1.9$
in notations of~\cite{Liu_Moriyama_11}.

\begin{figure}[ht]
\includegraphics[width=0.5\textwidth]{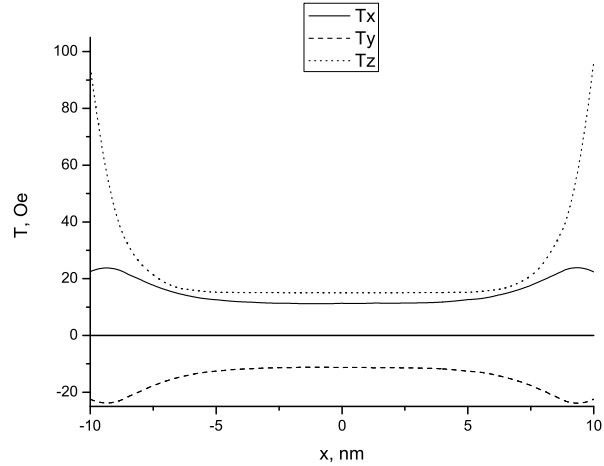} 
\caption{\label{fig:T20MxMy} Effective torques for the case of $\vec U_M=(\cos\pi/4;\sin\pi/4;0)$ near
the interface. Averaged values over volume of $Co$ are $<T_x>=15\,Oe$, $<T_x>=-15\,Oe$,
$<T_z>=3\,Oe$.}
\end{figure} 

\begin{figure}[ht]
\includegraphics[width=0.5\textwidth]{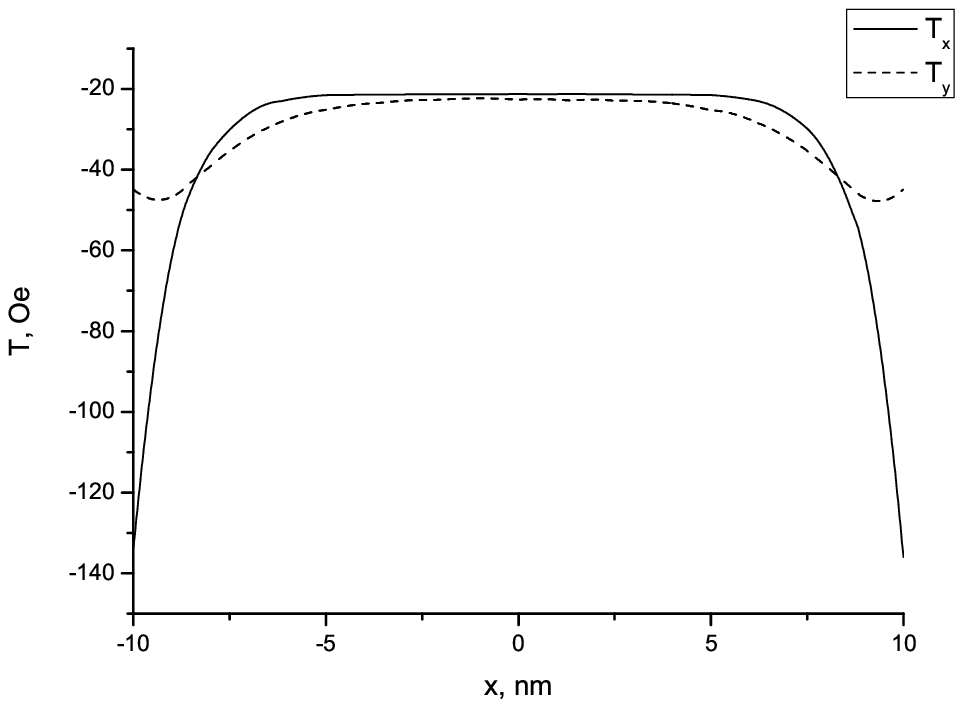} 
\caption{\label{fig:T20Mz} Effective torques near the interface in case of $M\parallel Oz$.
Averaged values over volume of $Co$ are $<T_x>=-4\,Oe$, $<T_y>=-30\,Oe$.}
\end{figure}

To investigate the possibility to manipulate with magnetization of small ferromagnetic dots under influence
of SHE current we have calculated the field induced by SHE in the small ferromagnetic dot situated on the
surface of paramagnetic metallic layer with SHE as shown in Fig.~\ref{fig:sys20x20}. The
direction of magnetization of the ferromagnetic dot forms the angle $\pi/4$ with $x$-axes, or it is parallel
or antiparallel to $y$ axes. In the case $\vec M\parallel y$ only $y$-component of induced fields not zero and its average
value does not change under inversion of magnetization direction along $y$ axes. This field is much stronger
than Oersted field which is about $5$~Oe. Besides that field does not produce any torque but defines the most
energetically favorable direction of magnetization. When this direction is not collinear with $y$-axes, the
SHE produces not only $y$-component of spin accumulation but all three $xyz$ components of spin accumulation due to the precession of spin
accumulation vector in exchange field of ferromagnet.

In fig.\ref{fig:T20MxMy} the dependence on $x$ coordinate of three components
of spin transfer torque, produced by spin accumulation due to the SHE are shown. The torque acting
on magnetisation is defined as
$\left[\vec M\times \vec H_{\mathrm{SHE}}\right]$, and in LLG equation it has to be multiplied by the
gyromagnetic ratio $\gamma$. For the case $U_m=(\cos\pi/4; \sin\pi/4; 0)$
the $xy$ part of torque vector $\vec T=(T_x; T_y; 0)$
is lying in $xy$ plane perpendicular to $\vec U_m$ and may by considered like an
additional damping or antidamping term in LLG equation.
The torque due to Oersted field has only $z$-component.

For the case of $\vec M\parallel Oz$, $T_x$ and $T_y$ torque components value close to the interface are shown in fig.\ref{fig:T20Mz}. 
These torques try to reorient magnetization of FM layer onto $xy$ plane and $T_y$ component may be considered like damping or antidamping
term. We checked the results taking the spin Hall conductivity equal to zero and obtained that all torques
besides one produced by Oersted field of the current vanish.

\section{Conclusions}
We have shown that SHE may represent a powerful tool for the manipulation with magnetization of small ferromagnetic
metal dots situated on the surface of the thin paramagnetic metal layer with large value of SHE conductivity.
Especially important feature of spin transfer torques created by spin accumulation due to influence of SHE is that these
torques have component similar to damping or antidamping spin torque produced by current in non-collinear
magnetic multilayers.

\section{Acknowledgments}
This works has been supported by Russian Fund of Fundamental Research, ERC Advanced Grant "HYMAGINE" and by French National Research Agency Project ANR-10-BLANC "SPINHALL".
\bibliography{sh}

\begin{thebibliography}{16}
\expandafter\ifx\csname natexlab\endcsname\relax\def\natexlab#1{#1}\fi
\expandafter\ifx\csname bibnamefont\endcsname\relax
  \def\bibnamefont#1{#1}\fi
\expandafter\ifx\csname bibfnamefont\endcsname\relax
  \def\bibfnamefont#1{#1}\fi
\expandafter\ifx\csname citenamefont\endcsname\relax
  \def\citenamefont#1{#1}\fi
\expandafter\ifx\csname url\endcsname\relax
  \def\url#1{\texttt{#1}}\fi
\expandafter\ifx\csname urlprefix\endcsname\relax\def\urlprefix{URL }\fi
\providecommand{\bibinfo}[2]{#2}
\providecommand{\eprint}[2][]{\url{#2}}

\bibitem[{\citenamefont{D'yakonov and Perel'}(1971)}]{Dyakonov_Perel_71_1}
\bibinfo{author}{\bibfnamefont{M.~I.} \bibnamefont{D'yakonov}}
  \bibnamefont{and} \bibinfo{author}{\bibfnamefont{V.~I.}
  \bibnamefont{Perel'}}, \bibinfo{journal}{JETP} \textbf{\bibinfo{volume}{13}},
  \bibinfo{pages}{467} (\bibinfo{year}{1971}).

\bibitem[{\citenamefont{Dyakonov and Perel}(1971)}]{Dyakonov_Perel_71_2}
\bibinfo{author}{\bibfnamefont{M.~I.} \bibnamefont{Dyakonov}} \bibnamefont{and}
  \bibinfo{author}{\bibfnamefont{V.~I.} \bibnamefont{Perel}},
  \bibinfo{journal}{Phys. Lett. A} \textbf{\bibinfo{volume}{35}},
  \bibinfo{pages}{459} (\bibinfo{year}{1971}).

\bibitem[{\citenamefont{Fert et~al.}(1981)\citenamefont{Fert, Friederich, and
  Hamzic}}]{Fert1981231}
\bibinfo{author}{\bibfnamefont{A.}~\bibnamefont{Fert}},
  \bibinfo{author}{\bibfnamefont{A.}~\bibnamefont{Friederich}},
  \bibnamefont{and} \bibinfo{author}{\bibfnamefont{A.}~\bibnamefont{Hamzic}},
  \bibinfo{journal}{Journal of Magnetism and Magnetic Materials}
  \textbf{\bibinfo{volume}{24}}, \bibinfo{pages}{231 } (\bibinfo{year}{1981}).

\bibitem[{\citenamefont{Hirsch}(1999)}]{Hirsch_99}
\bibinfo{author}{\bibfnamefont{J.~E.} \bibnamefont{Hirsch}},
  \bibinfo{journal}{Phys. Rev. Lett.} \textbf{\bibinfo{volume}{83}},
  \bibinfo{pages}{1834} (\bibinfo{year}{1999}).

\bibitem[{\citenamefont{Zhang}(2000)}]{Zhang_SH_2000}
\bibinfo{author}{\bibfnamefont{S.}~\bibnamefont{Zhang}},
  \bibinfo{journal}{Phys. Rev. Lett.} \textbf{\bibinfo{volume}{85}},
  \bibinfo{pages}{393} (\bibinfo{year}{2000}).

\bibitem[{\citenamefont{Shchelushkin and
  Brataas}(2005{\natexlab{a}})}]{Shchelushkin_Brataas_2005_1}
\bibinfo{author}{\bibfnamefont{R.~V.} \bibnamefont{Shchelushkin}}
  \bibnamefont{and} \bibinfo{author}{\bibfnamefont{A.}~\bibnamefont{Brataas}},
  \bibinfo{journal}{Phys. Rev. B} \textbf{\bibinfo{volume}{71}},
  \bibinfo{pages}{045123} (\bibinfo{year}{2005}{\natexlab{a}}).

\bibitem[{\citenamefont{Shchelushkin and
  Brataas}(2005{\natexlab{b}})}]{Shchelushkin_Brataas_2005_2}
\bibinfo{author}{\bibfnamefont{R.~V.} \bibnamefont{Shchelushkin}}
  \bibnamefont{and} \bibinfo{author}{\bibfnamefont{A.}~\bibnamefont{Brataas}},
  \bibinfo{journal}{Phys. Rev. B} \textbf{\bibinfo{volume}{72}},
  \bibinfo{pages}{073110} (\bibinfo{year}{2005}{\natexlab{b}}).

\bibitem[{\citenamefont{Valet and Fert}(1993)}]{Valet_Fert_1993}
\bibinfo{author}{\bibfnamefont{T.}~\bibnamefont{Valet}} \bibnamefont{and}
  \bibinfo{author}{\bibfnamefont{A.}~\bibnamefont{Fert}},
  \bibinfo{journal}{Phys. Rev. B} \textbf{\bibinfo{volume}{48}},
  \bibinfo{pages}{7099} (\bibinfo{year}{1993}).

\bibitem[{\citenamefont{Zhang et~al.}(2002)\citenamefont{Zhang, Levy, and
  Fert}}]{Zhang_Levy_Fert_02}
\bibinfo{author}{\bibfnamefont{S.}~\bibnamefont{Zhang}},
  \bibinfo{author}{\bibfnamefont{P.~M.} \bibnamefont{Levy}}, \bibnamefont{and}
  \bibinfo{author}{\bibfnamefont{A.}~\bibnamefont{Fert}},
  \bibinfo{journal}{Phys. Rev. Lett.} \textbf{\bibinfo{volume}{88}},
  \bibinfo{pages}{236601} (\bibinfo{year}{2002}).

\bibitem[{\citenamefont{Strelkov et~al.}(2010)\citenamefont{Strelkov, Vedyayev,
  Gusakova, Buda-Prejbeanu, Chshiev, Amara, Vaysset, and
  Dieny}}]{Strelkov_IEEE_10}
\bibinfo{author}{\bibfnamefont{N.}~\bibnamefont{Strelkov}},
  \bibinfo{author}{\bibfnamefont{A.}~\bibnamefont{Vedyayev}},
  \bibinfo{author}{\bibfnamefont{D.}~\bibnamefont{Gusakova}},
  \bibinfo{author}{\bibfnamefont{L.~D.} \bibnamefont{Buda-Prejbeanu}},
  \bibinfo{author}{\bibfnamefont{M.}~\bibnamefont{Chshiev}},
  \bibinfo{author}{\bibfnamefont{S.}~\bibnamefont{Amara}},
  \bibinfo{author}{\bibfnamefont{A.}~\bibnamefont{Vaysset}}, \bibnamefont{and}
  \bibinfo{author}{\bibfnamefont{B.}~\bibnamefont{Dieny}},
  \bibinfo{journal}{Magnetics Letters, IEEE} \textbf{\bibinfo{volume}{1}},
  \bibinfo{pages}{3000304} (\bibinfo{year}{2010}).

\bibitem[{\citenamefont{Strelkov et~al.}(2011)\citenamefont{Strelkov, Vedyayev,
  Ryzhanova, Gusakova, Buda-Prejbeanu, Chshiev, Amara, de~Mestier, Baraduc, and
  Dieny}}]{Strelkov_PRB_11}
\bibinfo{author}{\bibfnamefont{N.}~\bibnamefont{Strelkov}},
  \bibinfo{author}{\bibfnamefont{A.}~\bibnamefont{Vedyayev}},
  \bibinfo{author}{\bibfnamefont{N.}~\bibnamefont{Ryzhanova}},
  \bibinfo{author}{\bibfnamefont{D.}~\bibnamefont{Gusakova}},
  \bibinfo{author}{\bibfnamefont{L.~D.} \bibnamefont{Buda-Prejbeanu}},
  \bibinfo{author}{\bibfnamefont{M.}~\bibnamefont{Chshiev}},
  \bibinfo{author}{\bibfnamefont{S.}~\bibnamefont{Amara}},
  \bibinfo{author}{\bibfnamefont{N.}~\bibnamefont{de~Mestier}},
  \bibinfo{author}{\bibfnamefont{C.}~\bibnamefont{Baraduc}}, \bibnamefont{and}
  \bibinfo{author}{\bibfnamefont{B.}~\bibnamefont{Dieny}},
  \bibinfo{journal}{Phys. Rev. B} \textbf{\bibinfo{volume}{84}},
  \bibinfo{pages}{024416} (\bibinfo{year}{2011}).

\bibitem[{\citenamefont{Fert and Levy}(2011)}]{Fert_Levy_2011}
\bibinfo{author}{\bibfnamefont{A.}~\bibnamefont{Fert}} \bibnamefont{and}
  \bibinfo{author}{\bibfnamefont{P.~M.} \bibnamefont{Levy}},
  \bibinfo{journal}{Phys. Rev. Lett.} \textbf{\bibinfo{volume}{106}},
  \bibinfo{pages}{157208} (\bibinfo{year}{2011}).

\bibitem[{\citenamefont{Takeshi et~al.}(2008)\citenamefont{Takeshi, Yu, Seiji,
  Saburo, Hiroshi, Sadamichi, Junsaku, and Takanashi}}]{Takeshi_nmat_08}
\bibinfo{author}{\bibfnamefont{S.}~\bibnamefont{Takeshi}},
  \bibinfo{author}{\bibfnamefont{H.}~\bibnamefont{Yu}},
  \bibinfo{author}{\bibfnamefont{M.}~\bibnamefont{Seiji}},
  \bibinfo{author}{\bibfnamefont{T.}~\bibnamefont{Saburo}},
  \bibinfo{author}{\bibfnamefont{I.}~\bibnamefont{Hiroshi}},
  \bibinfo{author}{\bibfnamefont{M.}~\bibnamefont{Sadamichi}},
  \bibinfo{author}{\bibfnamefont{N.}~\bibnamefont{Junsaku}}, \bibnamefont{and}
  \bibinfo{author}{\bibfnamefont{K.}~\bibnamefont{Takanashi}},
  \bibinfo{journal}{Nature Materials} \textbf{\bibinfo{volume}{7}},
  \bibinfo{pages}{125} (\bibinfo{year}{2008}).

\bibitem[{\citenamefont{Takeshi}()}]{Takeshi_private_1}
\bibinfo{author}{\bibfnamefont{S.}~\bibnamefont{Takeshi}},
  \bibinfo{note}{private communication}.

\bibitem[{\citenamefont{Niimi et~al.}(2011)\citenamefont{Niimi, Morota, Wei,
  Deranlot, Basletic, Hamzic, Fert, and Otani}}]{Niimi_11}
\bibinfo{author}{\bibfnamefont{Y.}~\bibnamefont{Niimi}},
  \bibinfo{author}{\bibfnamefont{M.}~\bibnamefont{Morota}},
  \bibinfo{author}{\bibfnamefont{D.~H.} \bibnamefont{Wei}},
  \bibinfo{author}{\bibfnamefont{C.}~\bibnamefont{Deranlot}},
  \bibinfo{author}{\bibfnamefont{M.}~\bibnamefont{Basletic}},
  \bibinfo{author}{\bibfnamefont{A.}~\bibnamefont{Hamzic}},
  \bibinfo{author}{\bibfnamefont{A.}~\bibnamefont{Fert}}, \bibnamefont{and}
  \bibinfo{author}{\bibfnamefont{Y.}~\bibnamefont{Otani}},
  \bibinfo{journal}{Phys. Rev. Lett.} \textbf{\bibinfo{volume}{106}},
  \bibinfo{pages}{126601} (\bibinfo{year}{2011}).

\bibitem[{\citenamefont{Liu et~al.}(2011)\citenamefont{Liu, Moriyama, Ralph,
  and Buhrman}}]{Liu_Moriyama_11}
\bibinfo{author}{\bibfnamefont{L.}~\bibnamefont{Liu}},
  \bibinfo{author}{\bibfnamefont{T.}~\bibnamefont{Moriyama}},
  \bibinfo{author}{\bibfnamefont{D.~C.} \bibnamefont{Ralph}}, \bibnamefont{and}
  \bibinfo{author}{\bibfnamefont{R.~A.} \bibnamefont{Buhrman}},
  \bibinfo{journal}{Phys. Rev. Lett.} \textbf{\bibinfo{volume}{106}},
  \bibinfo{pages}{036601} (\bibinfo{year}{2011}).

\end{thebibliography}

\end{document}